\documentclass[twocolumn,showpacs,preprintnumbers,amsmath,amssymb]{revtex4}

\usepackage{graphicx}
\usepackage{dcolumn}
\usepackage{bm}

\begin{document}


\title{Single photon continuous variable quantum key distribution based on energy-time uncertainty relation}

\author{Bing Qi}
\affiliation{Department of Physics, University of Toronto, Toronto,
Ontario, M5S 1A7, Canada
}%

\date{\today}

\begin{abstract}
In previous quantum key distribution (QKD) protocols, information is
encoded on either the discrete-variable of single-photon signal or
continuous-variables of multi-photon signal. Here, we propose a new
QKD protocol by encoding information on continuous-variables of a
single photon. In this protocol, Alice randomly encodes her
information on either the central frequency of a narrowband single
photon pulse or the time-delay of a broadband single photon pulse,
while Bob randomly chooses to do either frequency measurement or
time measurement. The security of this protocol rests on the
energy-time uncertainty relation, which prevents Eve from
simultaneously determining both frequency and time information with
arbitrarily high resolution. In practice, this scheme may be more
robust against various channel noises, such as polarization and
phase fluctuations.
\end{abstract}

\maketitle


Unlike conventional cryptography, quantum key distribution (QKD)
provides unconditional security guaranteed by the fundamental laws
of quantum physics \cite{BB84,A91,Gisin02,securityproof}. To date,
both discrete-variable, single-photon QKD protocols (such as the
well-known BB84 protocol \cite{BB84})and continuous-variable,
multi-photon QKD protocols (such as QKD with squeezed
states\cite{Squeeze}) have been developed. In the standard BB84
protocol \cite{BB84,BB84more}, one of the legitimate users, Alice,
encodes information in a two-dimensional subspace (such as the
polarization state) of a single photon. The security is based on the
no-cloning theorem \cite{nonclone}. In contrast, in a continuous
variable QKD (CV-QKD), information is encoded on field quadratures
of either squeezed states or coherent states\cite{Squeeze,CV}(We
name this scheme as quadrature-coding CV-QKD to distinguish it from
the new protocol we will present). Quantum mechanically, the complex
field amplitudes correspond to a pair of non-commuting operators
$X_1$ and $X_2$. The security of this protocol follows from the
uncertainty relation for these two operators \cite{Squeeze}
\begin{equation}
\Delta X_1\Delta X_2\geq 1/4,
\end{equation}
It is thus impossible to measure both of them arbitrarily
accurately.

However, there are a few practical difficulties in telecom
fiber-based QKD. In both the phase-coding BB84 QKD system and the
quadrature-coding CV-QKD system, the quantum bit error rate (QBER)
is closely related to the interference visibility, which suffers
from polarization and phase instabilities induced by the optical
fiber. A Bidirectional auto-compensating structure has been
introduced to improve the performance of a practical QKD
system\cite{PP}. Unfortunately, this design also opens a potential
back-door for the eavesdropper (Eve) to launch various Trojan horse
attacks \cite{trojan}.

Normally, the frequency of a weak laser pulse won't change as it
propagates through fiber, and the temporal broadening can be well
controlled by employing dispersion-compensation techniques. This
inspires us to explore a frequency/time coding QKD protocol.

We remark that QKD protocols based on frequency-coding have been
studied previously\cite{previous1, previous2, previous3}. In
previous work\cite{previous2}: Alice represents bit 0 and bit 1 with
two distinguishable signal states $S_0$ (a single photon state with
frequency $\omega_0$) and $S_1$ (frequency $\omega_1$). She randomly
prepares either one of the two signal states or a control state
$S_c$, which is a superposition of $S_0$ and $S_1$. At Bob's side,
he randomly chooses to do one of the following three measurements:
$S_0$ measurement with a narrowband filter centered at frequency
$\omega_0$ and a single photon detector (SPD), $S_1$ measurement
with a $\omega_1$ filter and a SPD, or time measurement with a
time-resolving SPD (no filter). Alice and Bob uses signal states for
key distribution and control state for detecting Eve's attack. Note
in the cases when Alice prepares the control states and Bob does the
time measurements, a high visibility interference pattern is
expected. Any attack from Eve will unavoidably blur the interference
pattern and be caught.

We remark that in the above protocol\cite{previous2}, the density
matrix for the control state $\hat{\rho}_c$ is different from that
for the signal states $\hat{\rho}_s$ \cite{densitymatrix1}. In
principle, Eve can unambiguously distinguish them with a non-zero
probability\cite{attack}. We remark that to eliminate this
information, Alice can randomly prepare two types of control states,
$(|\omega_0\rangle+exp[i(\omega_1-\omega_0)t]|\omega_1\rangle)$ and
$(|\omega_0\rangle-exp[i(\omega_1-\omega_0)t]|\omega_1\rangle)$,with
equal probability. In this case,
$\hat{\rho}_s=\hat{\rho}_c=|\omega_0\rangle\langle\omega_0|+|\omega_1\rangle\langle\omega_1|$.

In this letter, we propose a single photon CV-QKD protocol: Alice
randomly encodes her information on either the central frequency or
the time-delay of a transform limited (TL) single photon
pulse\cite{sync}, while Bob randomly does frequency or time
measurement. The security of this protocol can be understood from
the well-known energy-time uncertainty relation $\Delta E\Delta
t\geq \hbar/4$. For a TL Gaussian pulse, it's easy to show that
\cite{TL}
\begin{equation}
\sigma_\omega\sigma_t=1,
\end{equation}
where $\sigma_\omega$ and $\sigma_t$ are the half width ($1/e$) of
the intensity spectrum and the temporal profile respectively. Eq.(2)
indicates that it's impossible to acquire both the frequency and the
time information of a single photon pulse with an arbitrarily high
resolution.

Compareing (1) with (2), we can see a high similarity between our
protocol and the squeezed states CV-QKD. We remark that the
energy-time uncertainty relation may be different fundamentally from
others, such as the position-momentum uncertainty relation, because
time is not an observable in non-relativistic quantum mechanics.
However, this fundamental question is outside the scope of our
current paper.

Fig.1 shows a diagram of our proposed QKD system. In Fig.1, Alice
randomly fires one of two single photon sources: $S_1$ produces
narrowband frequency tunable TL Gaussian pulses (bandwidth
$\sigma_{\omega1}$), while $S_2$ produces broadband time-delay
tunable TL Gaussian pulses(bandwidth $\sigma_{\omega2}$, central
frequency $\omega_0$), with $\sigma_{\omega1}<<\sigma_{\omega2}$.
The frequency tunable range of $S_1$ matches the spectrum of $S_2$,
while the time-delay tunable range of $S_2$ matches with the
temporal profile of $S_1$, as shown in Fig.2. A beam splitter $BS$
is employed to combine the outputs from $S_1$ and $S_2$ together. At
Bob's side, passively determined by a beam splitter, he can either
conduct time measurement with a high-speed time-resolving single
photon detector (TSPD), or frequency measurement with a dispersive
element (such as a dispersive grating) followed by a single photon
detector array (SPDA).
\begin{figure}[!t]\center
\resizebox{7.5cm}{!}{\includegraphics{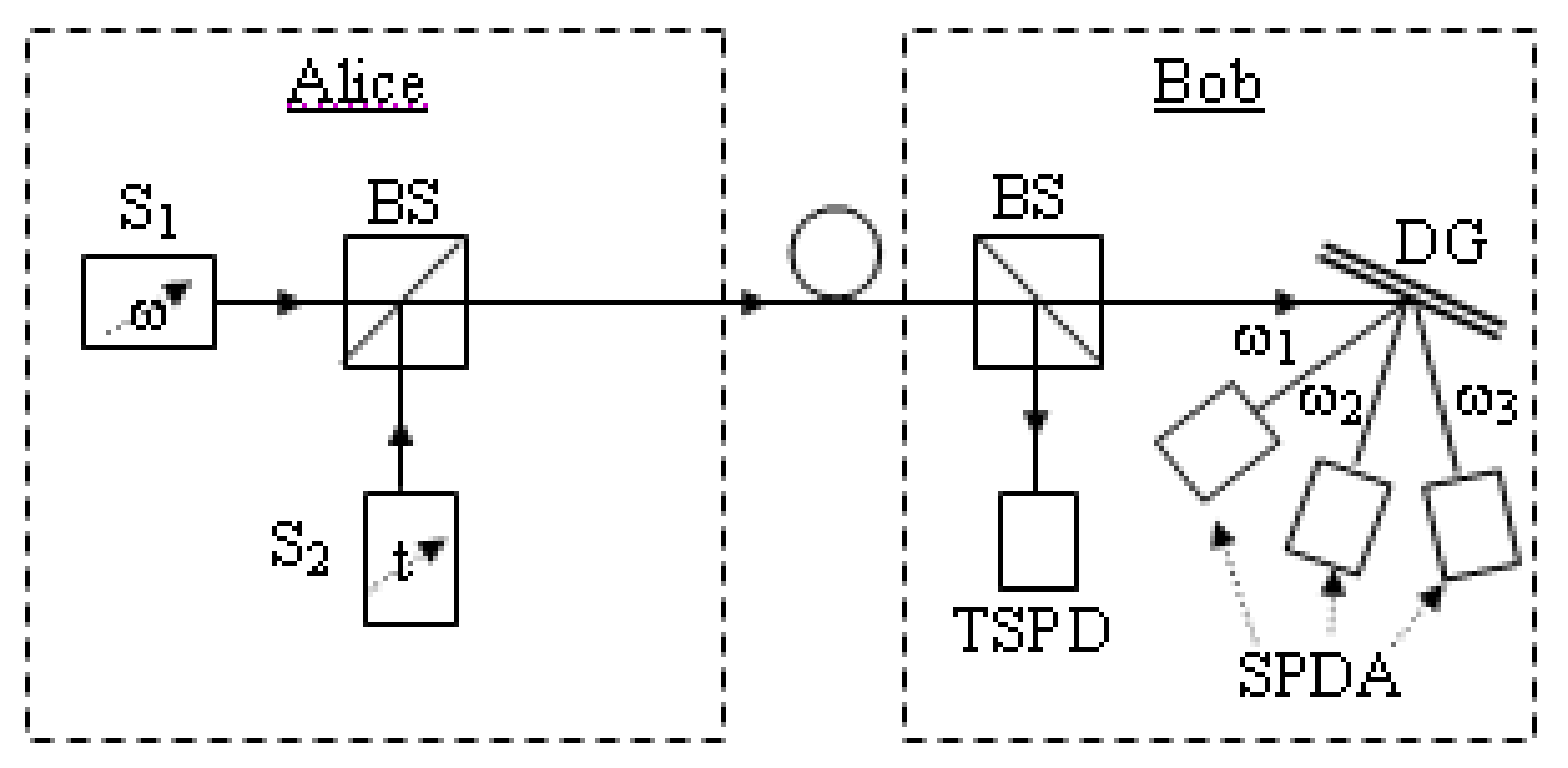}} \caption{Schematic
diagram of the proposed QKD system: $S_1$-narrowband frequency
tunable single photon source; $S_2$-broadband single photon source
with tunable time-delay; BS-beam splitter; DG-dispersive grating;
TSPD-time-resolving single photon detector; SPDA-single photon
detector array.}
\end{figure}
\begin{figure}[!t]\center
\resizebox{7.5cm}{!}{\includegraphics{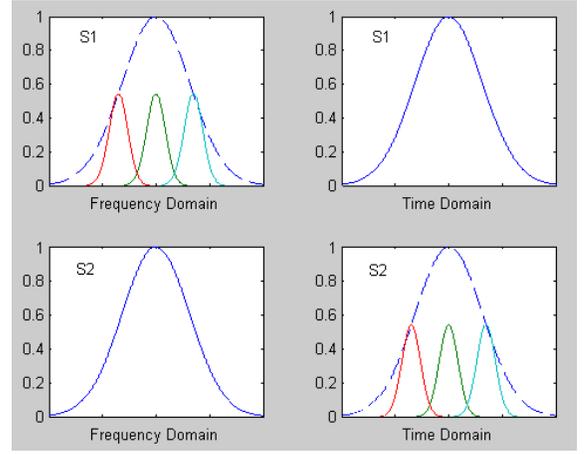}}
\caption{Illustration of the frequency (time) domain distributions
of Alice's single photon sources: top left-$S_1$ in frequency domain
(note the envelope matches with the spectrum of $S_2$); top
right-$S_1$ in time domain; bottom left-$S_2$ in frequency domain;
bottom right-$S_2$ in time domain(note the envelope matches with the
temporal profile of $S_1$).}
\end{figure}

Our QKD protocol runs as follows:

1.Alice generates a binary random number $a$. If $a=0$, she
generates another random number $b$ from the Gaussian distribution
$f_1(b)=(\pi\sigma_{\omega2}^2)^{-1/2}exp[-(b-\omega_0)^{2}/{\sigma_{\omega2}^{2}}]$;
then she sets the central frequency of $S_1$ to $b$ and fires it. If
$a=1$, Alice generates a random number $b$ from the Gaussian
distribution
$f_2(b)=(\pi)^{-1/2}\sigma_{\omega1}exp[-{\sigma_{\omega1}^{2}b^2}]$;
then she sets the time-delay of $S_2$ to $b$ and fires it.

2.Passively determined by a beam splitter, Bob either conducts time
measurement with a high-speed time-resolving single photon detector
(TSPD), or frequency measurement with a dispersive grating (DG)
followed by a single photon detector array (SPDA).

3.They repeat step 1 and step 2 many times.

4.Through an authenticated classical communication channel, they
post-select the cases when the quantum states prepared by Alice
match with the measurements conducted by Bob. After this step, Alice
and Bob share a set of correlated Gaussian variables, which are
called "key elements".

5.Alice and Bob can use the "sliced reconciliation" protocol
\cite{reconciliation} to transform the "key elements" into errorless
bit strings.

6.Alice and Bob can estimate the maximum information acquired by Eve
from the measured error rate and may use standard privacy
amplification protocol to distill out the final secure key.

The security of this protocol is based on: first, Eve can not
distinguish frequency-coding photons from time-coding photons;
secondly, Eve's ability to simultaneously determine both the
frequency information and the time information is constrained by the
uncertainty relation (2). It can be shown that in the asymptotic
case when $\sigma_{\omega1}\rightarrow0$, the density matrixes of
frequency-coding photons and time-coding photons are
identical\cite{densitymatrix2}:
\begin{equation}
\begin{aligned}
\hat{\rho}_1=\hat{\rho}_2=\int{(\pi\sigma_{\omega2}^2)^{-1/2}exp[-(\omega_1-\omega_0)^{2}/{\sigma_{\omega2}^{2}}]}\\
\times{\hat{a}^{+}(\omega_1){|0\rangle\langle0|}\hat{a}(\omega_1){d\omega_1}},
\end{aligned}
\end{equation}
Here, $\hat{a}^{+}(\omega)$ is the continuous-mode creation
operator. In practice, as long as
$\sigma_{\omega1}<<\sigma_{\omega2}$, the difference between the two
density matrixes is negligible.

To demonstrate the feasibility of this protocol, let's discuss the
basic requirements to the system's parameters. At Alice's side, the
spectral width of $S_1$ is $\sigma_{\omega1}$ with a continuous
tunable range of $\sigma_{\omega2}$, and the pulse width of $S_2$ is
$\sigma_{t2}=1/\sigma_{\omega2}$ with a continuous time-delay
tunable range of  $\sigma_{t1}=1/\sigma_{\omega1}$. At Bob's side,
the spectral and time resolutions of his measurement device are
$\delta_\omega$ and $\delta_t$ respectively. Further more, Alice and
Bob agree to slice each "key element" into bins of size $\Delta_t$
for a time-coding signal or $\Delta_\omega$ for a frequency-coding
signal. Here, we assume that
$\sigma_{\omega2}\gg\Delta_\omega\gg\delta_\omega\gg\sigma_{\omega1}$
and $\sigma_{t1}\gg\Delta_t\gg\delta_t\gg\sigma_{t2}$ as shown in
Fig.3a (for the case of time-coding).
\begin{figure}[!t]\center
\resizebox{7.5cm}{!}{\includegraphics{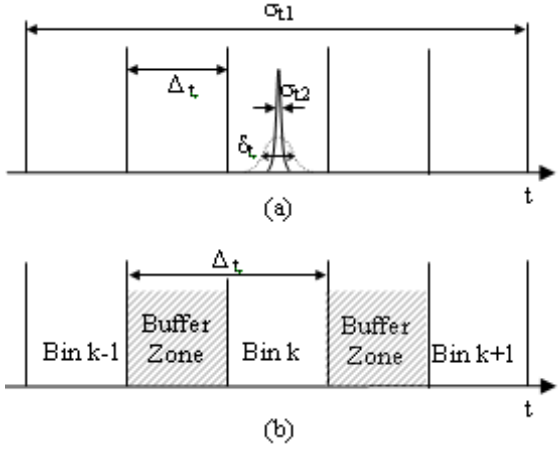}} \caption{(a) Slice
method without "buffer zone" (for time-coding signal) (b) Slice
method with "buffer zone" between neighbored bins (for time-coding
signal)}
\end{figure}

The error probability of time-coding signals can be estimated as
follows: Alice sends out photons at time
$t_c\in[-\Delta_t/2,\Delta_t/2]$, with a uniform distribution (for a
specific time bin). The intensity distribution in time domain
measured by Bob is
\begin{equation}
I(t)=(\pi\delta_t^2)^{-1/2}exp[-(t-t_c)^{2}/{\delta_t^2}],
\end{equation}
The probability that Bob's measurement results lies in the interval
$[-\Delta_t/2,\Delta_t/2]$ is
\begin{equation}
P_{\Delta
t}=(1/{\Delta_t})\int_{-{\Delta_t}/2}^{{\Delta_t}/2}\int_{-{\Delta_t}/2}^{{\Delta_t}/2}{I(t)dtdt_c},
\end{equation}

The error probability $P_e=(1-P_{\Delta_t})$ can be derived as
\begin{equation}
P_e=1-(1/{\sqrt{\pi}S_t})[-1+\sqrt{\pi}S_t erf(S_t)+exp(-S_t^2)],
\end{equation}
Here, $S_t=\Delta_t/\delta_t$, and $erf(x)$ is the error function
which is given by
\begin{equation}
erf(x)=1/\sqrt{\pi}\int_{0}^{x}{exp(-t^2)dt},
\end{equation}
In the case of frequency-coding, we can define
$S_\omega=\Delta_\omega/\delta_\omega$ and derive a similar formula.

From (6), to get a low error rate, $S_t$ ($S_\omega$)has to be large
enough. This can be achieved by either increasing the "bin size"
$\Delta_t$($\Delta_\omega$) or improving Bob's measurement
resolution $\delta_t$($\delta_\omega$). On the other hand, to
guaranty the security of this protocol, the bin sizes have to
satisfy the condition of $\Delta_\omega\Delta_t<1$. These conditions
set basic requirements for Bob's measurement resolutions $\delta_t$
and $\delta_\omega$\cite{measurement}. For example, from (6), the
error probability $P_e$ is about $0.056$ at $S_t=10$ for time-coding
or $S_\omega=10$ for frequency-coding. The requirement for Bob's
apparatus will be $\delta_t\delta_\omega<0.01$.

We remark that Alice/Bob can improve the slice method by adding in a
"buffer zone" between neighboring bins, as shown in Fig.3b (for the
case of time-encoding). In this scenario, Alice and Bob slices their
Gaussian variables into discrete bins with a "buffer zone" between
neighbored ones. During the classical communication stage, Alice
announces which pulses are prepared in the buffer zone, and they
just drop these results. In the case when Alice encodes her
information in the "bin zone", there are three possible outputs from
Bob's measurement: Bob detects a signal in the same bin with
probability $P_r$; Bob detects a signal in "buffer Zones" with
probability $P_b$ (this is an inconclusive result and they just drop
it); or Bob detects a signal in other bins with probability
$P_w=1-P_r-P_b$. Define $\Delta_t$ as the total size of one bin plus
one buffer zone, and assume that the bin size is equal to buffer
zone size, then $P_r$, $P_b$ can be derived as
\begin{equation}
P_r=({\Delta_t}/2)^{-1}\int_{-{\Delta_t}/4}^{{\Delta_t}/4}\int_{-{\Delta_t}/4}^{{\Delta_t}/4}{I(t)dtdt_c},
\end{equation}
\begin{equation}
P_b=({\Delta_t}/2)^{-1}\int_{-{\Delta_t}/4}^{{\Delta_t}/4}\int_{-3{\Delta_t}/4}^{3{\Delta_t}/4}{I(t)dtdt_c}-P_r,
\end{equation}
The error probability is defined as
\begin{equation}
P_e=P_w/(P_r+P_w),
\end{equation}

For $S_t=3$, the error probability $P_e$ calculated from (10) is
about $0.0038$ while the probability of getting an inconclusive
results is about $0.36$. Compared with the slice method without a
"buffer-zone", the error probability drops by about one order, while
the efficiency also drops by a factor of $0.5\times0.64=0.32$( the
$0.5$ factor is because that half of the time Alice sends out
signals in the buffer zone, and the $0.64$ factor is due to the
inconclusive probability in Bob's measurement). The requirement for
Bob's apparatus is $\delta_t\delta_\omega<0.1$, which can be
satisfied with today's technology\cite{resolution}. In practice, the
buffer size could be optimized to achieve the maximum secure key
rate.

The QKD protocol proposed here has several advantages: First of all,
a QKD system based on freqnency/time coding is intrinsically
insensitive to the polarization and phase fluctuations. This could
improve the stability of a fiber-based one-way QKD system
dramatically. Secondly, unlike the squeezed states QKD, our protocol
can be implemented with commercial laser sources. TL laser pulses
with different bandwidths can be easily prepared with commercial
products and they can go through long fibers with negligible
distortions (For example, in \cite{dispersion}, after going through
a $50km$ fiber, a $460fs$ pulse was slightly broaden to $470fs$.
This is orders lower than the time resolution of today's
SPD).Thirdly, compared with previous frequency-coding
protocol\cite{previous2}, our system could achieve a higher key rate
by using a large alphabet.

We remark that instead of using single photon sources (reliable
perfect single photon sources are far from practical), highly
attenuated laser sources could be employed to implement our
protocol. In this case, to guard against the PNS (photon number
splitting) attack, the newly developed decoy state idea can be
adopted\cite{decoy}.

In conclusion, we propose a new QKD protocol by encoding information
on continuous-variables of single photon signals. The security of
this protocol rests on the energy-time uncertainty relation, which
prevents Eve from simultaneously determine both frequency and time
information with arbitrarily high resolution. It will be interesting
to see whether a QKD protocol based on the energy-time uncertainty
relation is equivalent to the one based on uncertainty relation for
two non-commuting operators (such as the squeezed states QKD). In
practice, this scheme may be more robust against various channel
noises, such as polarization and phase fluctuations.

The author is very grateful to Hoi-Kwong Lo for his supports and
helpful comments. The author also thanks Xiong-Feng Ma, Yi Zhao, Ben
Fortescue and Chi-Hang Fred Fung for helpful discussions.


\begin{thebibliography}{00}
\bibitem{densitymatrix1}
The density matrix of the signal states is
$\hat{\rho}_s=|\omega_0\rangle\langle\omega_0|+|\omega_1\rangle\langle\omega_1|$.
The control state can be descried by
$|\omega_0\rangle+exp[i(\omega_1-\omega_0)t]|\omega_1\rangle$ with a
density matrix
$\hat{\rho}_c=|\omega_0\rangle\langle\omega_0|+|\omega_1\rangle\langle\omega_1|
+exp[-i(\omega_1-\omega_0)t]|\omega_0\rangle\langle\omega_1|
+exp[i(\omega_1-\omega_0)t]|\omega_1\rangle\langle\omega_0|$. In
general, $\hat{\rho}_s \neq \hat{\rho}_c$.
\bibitem{attack}
Eve could employ a high speed switch which is opened at
$t_{min}$(when the amplitude of control state is expected to be
zero) for a time-duration of $\Delta T\ll1/(\omega_1-\omega_0)$. The
signal states can pass through this switch with a much higher
probability than the control state.
\bibitem{sync}
In practice, for time-coding, the information can be encoded on the
time interval between the signal pulse and a synchronization pulse.
\bibitem{densitymatrix2}
The continuous-mode photon wave-packet creation operator can be
defined as
$\hat{a}_\xi^{+}=\int{d\omega\xi(\omega)\hat{a}^{+}(\omega)}$. Here,
$\xi(\omega)$ is the spectral amplitude.  For a Gaussian pulse with
a central frequency of $\omega_0$,
$\xi(\omega)=(\pi\sigma_\omega^2)^{-1/4}exp[-i(\omega_0-\omega)t_0-(\omega-\omega_0)^{2}/{2\sigma_\omega^{2}}]$,
where $t_0$ is the time when the peak of the pulse passes the
coordinate origin. For $S_1$, $t_0=0$ and the central frequency
$\omega_1$ is chosen from
$f_1(\omega_1)=(\pi\sigma_{\omega2}^2)^{-1/2}exp[-(\omega_1-\omega_0)^{2}/{\sigma_{\omega2}^{2}}]$.
So,the density matrix is
$\hat{\rho}_1=\int{f_1(\omega_1)\hat{a}_{\xi1}^{+}{|0\rangle\langle0|}\hat{a}_{\xi1}d\omega_1}
=\int\int\int(\pi\sigma_{\omega2}^2)^{-1/2}exp[-(\omega_1-\omega_0)^{2}/{\sigma_{\omega2}^{2}}]
(\pi\sigma_{\omega1}^2)^{-1/4}exp[-(\omega_2-\omega_1)^{2}/{2\sigma_{\omega1}^{2}}]
(\pi\sigma_{\omega1}^2)^{-1/4}\times{exp[-(\omega_3-\omega_1)^{2}/{2\sigma_{\omega1}^{2}}]
\hat{a}^{+}(\omega_2){|0\rangle\langle0|}\hat{a}(\omega_3){d\omega_1}{d\omega_2}{d\omega_3}}$.
Similarly, we can derive the density matrix for $S_2$ is
$\hat{\rho}_2=\int{f_2(t)\hat{a}_{\xi2}^{+}{|0\rangle\langle0|}\hat{a}_{\xi2}dt}
=\int\int\int(\pi)^{-1/2}\sigma_{\omega1}exp[-{\sigma_{\omega1}^{2}t^2}]
(\pi\sigma_{\omega2}^2)^{-1/4}exp[-i(\omega_0-\omega_1)t-(\omega_1-\omega_0)^{2}/{2\sigma_{\omega2}^{2}}]
(\pi\sigma_{\omega2}^2)^{-1/4}exp[i(\omega_0-\omega_2)t-(\omega_2-\omega_0)^{2}/{2\sigma_{\omega2}^{2}}]
\hat{a}^{+}(\omega_1){|0\rangle\langle0|}\hat{a}(\omega_2){d\omega_1}{d\omega_2}{dt}
$ In the asymptotic case when $\sigma_{\omega1}\rightarrow0$, We
have
$\hat{\rho}_1=\hat{\rho}_2=\int{(\pi\sigma_{\omega2}^2)^{-1/2}exp[-(\omega_1-\omega_0)^{2}/{\sigma_{\omega2}^{2}}]}
\times{\hat{a}^{+}(\omega_1){|0\rangle\langle0|}\hat{a}(\omega_1){d\omega_1}}$.
\bibitem{measurement}
In practice, the resolutions of Alice's devices for state
preparetion are much higher than the resolutions of Bob's
measurement devices.
\bibitem{resolution}
Today's SPD has a time resolution in the order of
$\delta_t\sim10^{-10}S$, while the spectral resolution of a F-P
interferometer can be $10^7Hz$. This give us
$\delta_t\delta_\omega\sim0.001$.
\end{thebibliography}
\end{document}